\documentclass[10pt,letterpaper]{article}
\usepackage[top=0.85in,left=2.75in,footskip=0.75in,marginparwidth=2in]{geometry}

\usepackage[utf8]{inputenc}

\usepackage{cite}

\usepackage{nameref,hyperref}

\usepackage[right]{lineno}

\usepackage{microtype}
\DisableLigatures[f]{encoding = *, family = * }

\raggedright
\usepackage{changepage}

\usepackage[aboveskip=1pt,labelfont=bf,labelsep=period,singlelinecheck=off]{caption}

\makeatletter
\renewcommand{\@biblabel}[1]{\quad#1.}
\makeatother

\usepackage{lastpage,fancyhdr,graphicx}
\usepackage{epstopdf}
\fancyheadoffset[L]{2.25in}
\fancyfootoffset[L]{2.25in}

\usepackage{color}

\definecolor{Gray}{gray}{.25}

\usepackage{graphicx}

\usepackage{sidecap}

\usepackage{wrapfig}
\usepackage[pscoord]{eso-pic}
\usepackage[fulladjust]{marginnote}
\reversemarginpar

\begin{document}
\vspace*{0.35in}

\begin{flushleft}
{\Large
\textbf\newline{Power Law Rheology of Folded Protein Hydrogels}
}
\newline
\\
Anders Aufderhorst-Roberts$^{1,2}$
Sophie Cussons$^{3,4}$
David J. Brockwell,$^{3,4}$
Lorna Dougan$^{2,3\ast}$

\bigskip
\bf{1} Centre for Materials Physics, Department of Physics,\\ University of Durham, Durham, DH1 3LE, UK\\
\bf{2} School of Physics and Astronomy, \\University of Leeds, Leeds, LS2 9JT, UK \\
\bf{3} Astbury Centre for Structural Molecular Biology, \\University of Leeds, Leeds, UK\\
\bf{4} School of Molecular and Cellular Biology, Faculty of Biological Sciences, \\University of Leeds, Leeds, LS2 9JT, UK \\
\bigskip
*l.dougan@leeds.ac.uk.

\end{flushleft}

\section*{Abstract}
Folded protein hydrogels are prime candidates as tuneable biomaterials but it is unclear to what extent their mechanical properties have mesoscopic, as opposed to molecular origins. To address this, we probe hydrogels of the muscle-derived protein $I27_5$, using a multimodal rheology approach. Across multiple protocols, the hydrogels consistently exhibit power-law viscoelasticity in the linear viscoelastic regime with an exponent $\beta = 0.03$, suggesting a dense fractal meso-structure, with predicted fractal dimension $d_f = 2.48$. In the nonlinear viscoelastic regime, the hydrogel undergoes stiffening and energy dissipation, indicating simultaneous alignment and unfolding of the folded proteins. Remarkably, this behaviour is highly reversible, as the value of $\beta$, $d_f$ and the viscoelastic moduli return to their equilibrium value, even after multiple cycles of deformation. This highlights a previously unrevealed diversity of viscoelastic properties that originate on the mesoscopic scale.  These considerations are likely to be key to controlling the viscoelasticity of folded protein hydrogels. 


\section*{Introduction}

Hydrogels are hydrated three-dimensional networks with significant potential in applications that involve the repair and replacement of biological tissue.\cite{calo2015} A desired aspect of their design is that they mimic the material properties of the tissue that they are designed to replace. The ability to emulate biomechanical properties such as toughness, strain-stiffening and mechanical memory is therefore an important scientific challenge,  particularly for applications involving cell growth and proliferation.\cite{vedadghavami2017manufacturing} The design challenges to recreate these properties are significant since materials must be sufficiently soft to allow cell proliferation and sufficiently stiff to protect the material from large and sudden deformations.\cite{langer2004designing} In living matter, this trade-off is enabled by the unique mechanical properties of biopolymers and their ability to increase their stiffness when deformed and to accommodate large strains through force induced unfolding.\cite{burla2019mechanical} Discovering approaches that can recreate these properties in synthetic materials remains highly challenging.  

An increasingly promising approach to address this challenge is to construct chemically crosslinked hydrogels from folded proteins.\cite{Hughes2022,lv2012tandem,khoury2019chemical,wu2018rationally}  In nature, protein domains have precise structures which change their conformation under precise chemical and mechanical conditions. When engineered as polyproteins, their relatively small molecular size ($\sim$10-100kDa) distinguishes them from conventional biopolymers, however w hen crosslinked into hydrogels, polyproteins exhibit impressive mechanical properties such as high elasticity\cite{Lv2010} and resilience\cite{lv2012tandem}. These properties can be controlled through chemical stimuli\cite{kong2016metal,fu2019dynamic}, which suggests they may be used in applications involving shape change and shape programming.\cite{khoury2020cation,khoury2019chemical,bian2022engineering} Their use is further advantageous since the molecular mechanics of folded protein domains have been extensively interrogated using force spectroscopy\cite{petrosyan2021single,Rief1997,hoffmann2013towards} and can be correlated to bulk viscoelasticity .\cite{broedersz2014modeling,shmilovich2018modeling} 

All hydrogels are intrinsically cross-length scale materials; however, most studies of folded protein hydrogels have focused only on the molecular scale mechanics of the protein and on the bulk properties of the resulting hydrogel.  Recent findings from our group have shown that structural parameters at the meso-scale may also be of importance.  These include the intermediate structural hierarchy including cross-linking,\cite{Hanson2019,hanson2020network,hanson2021intermediate} which can be used to modulate the network growth and the fractal dimensions and force distribution of percolating clusters. \cite{Hughes2021,Hughes2020} This rich hierarchy of structural properties indicates that polyprotein hydrogels are likely to exhibit complex viscoelastic behaviours, which should manifest themselves in their mechanical response. Relating this mechanical response to the underlying structure can be achieved through shear rheology, since the time-dependent viscoelasticity of the hydrogel directly reflects the length scale of the hydrogel structure.\cite{Muthukumar1989} However, most studies of polyprotein hydrogels have instead used dynamic mechanical analysis,(DMA) which is instructive for quantifying bulk elastic properties such as toughness and fracture stress but less effective in quantifying viscoelastic and structural properties.  In comparison to shear rheology, DMA may also induce volumetric sample changes at high strains, which may frustrate accurate interpretation of the data.\cite{gasik2017viscoelastic} However, a multi-modal rheological characterisation of polyprotein hydrogels that relates time-dependent viscoelasticity to underlying structure has yet to be carried out. 

In this study, we seek to bridge the gap between the molecular and bulk mechanics of folded protein hydrogels, using shear rheological tests to probe their meso-scale mechanics.  We construct hydrogels using $I27_5$ polyproteins, so called because they comprise five concatenated $I27$ immunoglobulin domains of the muscle protein titin. Titin is an elastomeric component of the sarcomeres of striated muscle, where its $I27$ domains are able to unfold and refold to prevent overstretching.\cite{Improta1996} Reflecting this biological role, the $I27$ protein possesses mechanical stability whereas other domains in titin extend with low resistance. It also possesses a high degree of mechanical reversibility,\cite{best2003mechanical,chen2015dynamics} as well as being suitable for crosslinking into a hydrogel using photochemical approaches.\cite{da2017assessing} This makes it a putative robust and tough polyprotein and an excellent candidate for this study. We probe $I27_5$ hydrogels through frequency-dependent small amplitude oscillatory shear, stress relaxation and strain ramp experiments, to probe the viscoelastic properties at a range of timescales and within different regimes of deformation. By combining these different modes of rheology, we present a comprehensive rheological analysis of $I27_5$ hydrogels that infers their meso-scale structure,  reveals how this structure relates to viscoelasticity and explores how this viscoelasticity  determines the nonlinear and fracture mechanics of the hydrogel. 

\section*{Methods}
\subsubsection*{Polyprotein Purification}
The assembly of $I27_5$ polyproteins was performed using a PCR-based Golden Gate protocol with a modified pET14b, with all original BsaI sites removed, as the destination expression vector.\cite{Potapov2018} After verification of the DNA sequence, the resulting vector was transformed into BLR (DE3) pLysS \textit{E.coli} cells. 2 ml LB starter culture was used to inoculate 0.5 L autoinduction medium. 10 $\times$ 0.5 L cell cultures were incubated at 28$^\circ$C, 200 rpm for 24 h, for protein expression and then harvested and lysed. The protein was loaded on to 2 $\times$ 5 ml HisTrap FF columns (Cytiva) for Ni$^{2+}$ affinity chromatography. The column was then equilibrated in wash buffer (20 mM Tris, 300 mM NaCl, 10 mM imidazole, pH 8), before the protein was eluted with elution buffer (20 mM Tris, 300 mM NaCl, 500 mM imidazole, pH 8), in a ratio of 1:3 to wash buffer. The protein was further purified using a 5ml HiTrap DEAE Sepharose FF Ion Exchange chromatography column (Cytiva) for ion exchange purification. A gradient elution of 0-100$\%$ elution buffer (25 mM Sodium phosphate pH 7.4, 0-500 mM NaCl) was performed over 300 ml. The protein was then purified by size exclusion chromatography (HiLoad 26/600 Superdex 75 pg, Cytiva) and eluted in 25 mM Sodium Phosphate, pH 7.4. The purified protein was dialysed into MilliQ water and freeze-dried for storage at -20$^\circ$C.

\subsubsection*{Hydrogel Preparation}
$I27_5$ was re-suspended in 25 mM sodium phosphate,  pH = 7.4 and placed on a rotary wheel for a minimum of 1 h to enable complete dissolution of protein. The suspensions were then centrifuged for 10 minutes at 5000 RCF, retaining the supernatant to ensure that any insoluble protein was removed. Protein concentration was measured by absorption at 280  nm, using an extinction coefficient of 43824M$^{-1}$cm$^{-1}$. The final suspension was then diluted and mixed with an assembly buffer of Tris(2,2'-bipyridyl)-dichlororuthenium(II) hexahydrate (Ru(BiPy)$_3$) and sodium persulfate (NaPS) directly before use, with final concentrations of 0.38 mM $I27_5$, 100 $\mu$M Ru(BiPy)$_3$ and 10 mM NaPS.  Sample gelation was initiated photochemically through 60 seconds of in-situ illumination with a 460 nm light emitting diode as previously described\cite{AufderhorstRoberts2020}.   The illumination intensity was calibrated to 40 mW.cm$^{-2}$ at 452 nm. 

\subsubsection*{Shear Rheology}
Rheology measurements were performed using a stress-controlled rheometer (Kinexus Malvern Pro) equipped with a 10 mm steel plate and plate geometry. The lower plate of the rheometer was replaced with a custom-built photodiode module which allowed the photochemical crosslinking to be carried out in-situ. To prevent the evaporation, low viscosity (5 cSt) mineral oil was applied to the air-sample interface.  All measurements were performed at 25 $^\circ$C and 2 hours after crosslinking.  For all measurements, rheology data was captured over at least 3 measurements on independently prepared samples. 

\subsubsection*{Small Amplitude Oscillatory Measurements}
All small amplitude oscillatory measurements were carried out at a strain amplitude of $1\%$  - comfortably within the linear viscoelastic regime for all samples studied. Time dependent measurements were carried out at an angular frequency of 6.28 s$^-1$ unless otherwise stated. At high frequencies, artifacts from instrument inertia were avoided by discounting all data points for which the phase angle exceeded 150$^\circ$.

\subsubsection*{Cyclic Loading Measurements}
Cyclic loading experiments were performed to probe the mechanical response of the polyprotein hydrogel to large deformations and to examine the effect of multiple deformations on the hydrogel mechanics. Following crosslinking, the sample was subjected to a shear strain pulse of loading rate 0.01 s$^{-1}$.  After each pulse, the sample was left to recover for 23 min, during which time a single oscillatory frequency sweep was carried out to assess any permanent changes to material properties. The process was then repeated with incrementally increasing strain pulse magnitudes. 

\subsubsection*{Stress relaxation}
Stress relaxation experiments were carried out by applying a step strain to the sample for 10 mins. The resulting relaxation in stress over time was fitted to a range of different constitutive models using the Rheology Open Source (RHEOS) software.\cite{Kaplan2019,Bonfanti2020}  A constant applied strain  was typically attained within 2-3 s and any data from before this point were discarded from the analysis. 

\subsubsection*{Circular Dichroism}
Circular dichroism (CD) measurements were performed on an Applied Photophysics Chirascan CD spectropolarimeter using 0.01 mm path length cuvettes at a concentration of 0.5 mM. Samples were gelled inside the cuvette through photochemical crosslinking using identical buffers and the same protocol to that used in shear rheology experiments. Samples were also probed at higher dilution (5  $\mu$M), in which case the protein was probed in a phosphate buffer without crosslinking agents (25 mM sodium phosphate,  pH = 7.4) in a 1mm path length cuvette. Each measurement was averaged over 1 s and any data points for which the photomultiplier tube voltage (HV) exceeded 600 V were not included in the analysis.  A minimum of 4 independently prepared samples were used for each measurement to ensure reproducibility. Standard deviations did not exceed 10$\%$ of the mean value. 

\subsection*{Results}
\subsubsection*{I27 Protein Structure is Retained following Crosslinking}
\begin{figure*}[ht]\centering 
	\includegraphics[width=1\linewidth]{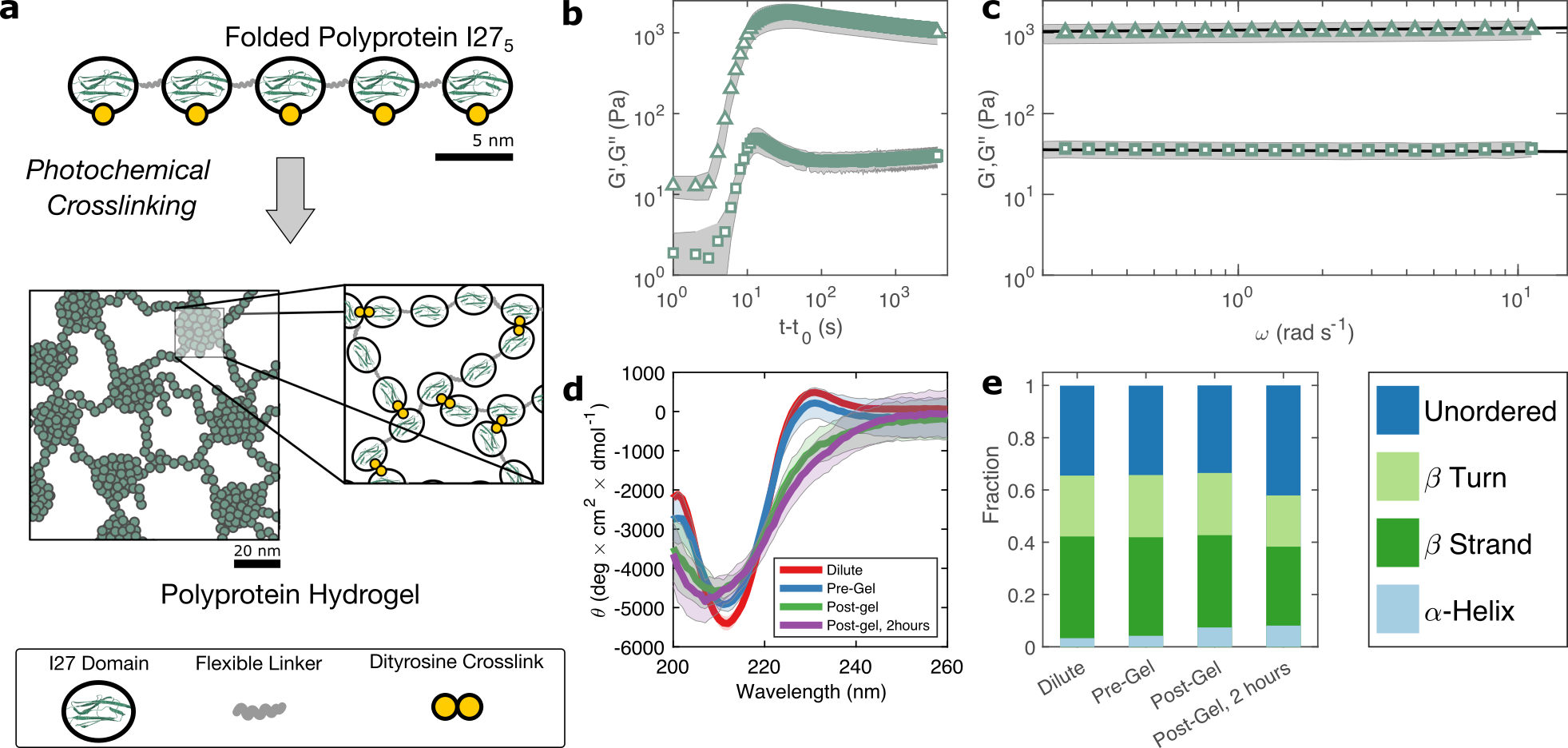}
	\caption{(a) The polyprotein used in this study comprises five concatenated I27 domains inter-spaced with flexible 'linker' regions. Illumination of polyprotein solutions with 490 nm blue light in the presence of crosslinking reagents leads to gelation through the formation of dityrosine bonds. (b) A sharp increase in G' (triangles) and G'' (squares) over a timescale of seconds indicates rapid crosslinking. (c) Within the linear viscoelastic regime, after 2 hours both G' and G''  exhibit a weak power law dependence on frequency (d) Circular dichroism spectra show a  $\beta$-sheet peak at 211  nm, with a slight peak shift and decrease in intensity after gelation. (e) Analytical fitting spectra using the CONTINLL reference set confirms the structure is predominantly  $\beta$-sheet, before and after gelation and indicates reduction in the fraction of folded protein of approximately 12$\%$, taking place at equivalent timescales to the decrease in G'/G''. Error bars in (b)-(d) (shaded regions) represent averaged measurements. }
	\label{FIG1}
\end{figure*}

The $I27_5$ polyprotein comprises five sequential repeats of the $I27$ as shown in figure \ref{FIG1} (a). The I27 polyproteins are assembled into hydrogels using the well-known ruthenium crosslinking strategy, \cite{fancy1999chemistry} which causes tyrosine residues on the surface of the polyproteins to radicalise and form dityrosine linkages under illumination of visible light and in the presence of sodium persulfate. Each polyprotein has five native tyrosine sites which are retained in the polyprotein design. The polyprotein also contains a hexahistidine repeat at its N-terminus and a dicysteine residue at its C-terminus, both of which are known to crosslink under the same conditions, albeit at lower efficiency.\cite{fancy2000scope} Therefore the polyprotein can be said to possess 7 potential crosslinking sites. The sodium persulfate that drives the crosslinking reaction is present at an approximately 4$\times$ stochiometric excess to the potential crosslink sites therefore we may assume a high crosslinking efficiency. Recent computational  work from our group has predicted that the coordination efficiency of pentameric polyproteins with a similar number of crosslinks to $I27_5$ is close to unity,\cite{hanson2021intermediate} therefore without making a priori assumptions about the meso-scale structure, it can already be assumed that the network is densely crosslinked, with each polyprotein being permanently crosslinked to multiple neighbouring polyprotein molecules. 

To probe this crosslinking process, we carry out small amplitude oscillatory shear measurements, probing the storage (G$'$) and loss (G$''$) moduli of the sample during \textit{in situ} illumination at time $t = t_0$. As plotted in figure \ref{FIG1}(b) there exists a short lag time before a sharp increase is observed in both G$'$ and G$''$. In a recent study of globular bovine serum albumin (BSA) hydrogels we showed that this lag time\cite{AufderhorstRoberts2020} is strongly indicative of  a two-step nucleation and growth process, in which diffusion-limited pre-gel clusters mature towards eventual percolation. Although the intensity of light and the stochiometric ratio are similar to this previous study, the observed lag time is significantly shorter ($\sim$ 3 s) for $I27_5$ than for BSA ($\sim 30$ s). 

The value of G' and G'' both increase sharply following the lag phase, reaching a peak of $1.7 \pm 0.6 \times 10^3$ Pa after approximately 30 s, before undergoing a steady decrease and reaching a final G' of $1.0 \pm 0.3 \times 10^3$ Pa, after approximately 1 hr.  To examine the possible molecular origins of this decrease in moduli, we carry out CD spectroscopy across each stage of the crosslinking process as shown in figure \ref{FIG1}(d). Before crosslinking the mean residue ellipticity spectrum indicates an anti-parallel  $\beta$-sheet structure which, in I27, is typically characterised by a negative band at 211  nm \cite{Wolny2014}. A slight positive band can also be seen close to 230  nm, which reflects exciton coupling between aromatic residues\cite{Batchelor2018,Sreerama2003}.  Compared with the spectra of dilute uncrosslinked $I27_5$, there is a close alignment ($<10\%$ difference in peak at 211  nm) indicating that the protein remains largely stable at the increased concentrations used for gelation, while the magnitude of the 211  nm band ($5.4 \pm 0.2 \times 10^3$ deg. cm$^2$. dmol$^{-1}$) is comparable to previous studies of $I27_5$. Following photochemical crosslinking, the peak intensity reduces and shifts to a lower wavelength, which may indicate a decrease in the folded protein fraction. Assuming that the 211  nm peak intensity is directly proportional to the fraction of folded protein, we estimate a reduction in folded fraction of $I27_5$ of 7$\%$. The 230  nm peak is also reduced, presumably due to the conversion of the  aromatic tyrosine residues into dityrosine residues during crosslinking.  Overall, this indicates a slight  reduction in the folded fraction of $I27_5$. To investigate this further, we deconvoluted the CD spectra using the CONTINLL parameter set in Dichroweb\cite{sreerama2000estimation}, using protein reference set 4 (optimised for 190 - 240  nm).  As shown in figure \ref{FIG1}(e),  table \ref{CD_table}, and figure \ref{SUPP-dichroweb} there is a slight reduction in $\beta$-sheet fraction  and $\beta$-turn fraction and an increase in the proportion of the protein that is unordered.  Taking this as being indicative of protein unfolding under gelation, we estimate a decrease in the fraction of folded protein of approximately 12$\%$, which takes place in the 2 hours following gelation, the same time over which a decrease in G' and G''  is observed. A similar reduction in folded fraction that coincides with a decrease in G' and G'' has previously been observed for other folded protein hydrogels studied in our group\cite{AufderhorstRoberts2020, Hughes2021, Hughes2020}, which suggests that it is a general feature of folded protein hydrogels, where it is caused by forced unfolding of weak protein domains on the \textit{molecular}-scale as the hydrogel network forms. 

\begin{table}[]
\begin{tabular}{p{0.13\textwidth}|p{0.087\textwidth}p{0.071\textwidth}p{0.05\textwidth}}
                                  & \textbf{$\beta$-Strands} & \textbf{$\beta$-Turns} & \textbf{Unordered}\\
                                  \hline
\textbf{Dilute (5 $\mu$M) }                             & 0.39             & 0.23           & 0.34                \\
\textbf{Pre-Crosslink}                   & 0.38             & 0.24           & 0.34         \\
\textbf{Post-Crosslink}                  & 0.35             & 0.24           & 0.34               \\
\textbf{Post-Gel, 2 hrs}                   & 0.30             & 0.20           & 0.42

\end{tabular}
\caption{ \label{CD_table} Secondary structure composition of I27$_5$ polyprotein hydrogels.  $\alpha$-helix content (not shown) is the result of deconvolution errors.  }
\end{table}

\subsubsection*{Frequency Sweeps Indicate a Fractal Network Structure}

Rheological frequency sweeps (figure \ref{FIG1}(c)) show that the equilibrated hydrogels are predominantly elastic in nature, with a storage modulus G' that is over an order of magnitude higher than the loss modulus G'' across all probed frequencies. Frequency sweeps reflect the distribution in timescales of the underlying relaxation mechanisms in the hydrogel.  For our I27$_5$ hydrogels we find that the frequency dependence of G' and G'' are well-described by weak power law relationships (solid lines in figure \ref{FIG1}(c)).   The physical origins of this power law viscoelasticity in hydrogels remain a matter of debate\cite{Aime2018} but they are commonly associated with fractal structure in a variety of soft and biological materials\cite{Aime2018,Takenaka2004,Dahesh2016,Muthukumar1989,Matsumoto1992,Patrcio2015,eleya2004scaling,Helmberger2014,jaishankar2014fractional,duval2010creep, Bremer1990}. In recent work from our group, we have shown that fractal structures are also present in folded protein hydrogels\cite{Hughes2020,Hughes2021} and that these structures arise from the diffusion- and rate-limited crosslinking of constituent folded proteins\cite{AufderhorstRoberts2020} and \textit{in situ} protein unfolding.\cite{Hughes2020,Hughes2021}

The emergence of a power law relaxation spectrum can be rationalised by considering that a fractal hydrogel network has a self-similar structure and consequentially will exhibit self-similar relaxation dynamics. \cite{Patrcio2015} Assuming that the excluded volume interactions are screened, this allows the fractal dimension $d_f$  of the hydrogel network to be directly calculated using the following previously derived\cite{Muthukumar1989} relation:

\begin{equation}
\beta = \frac{3(5-2d_f)}{2(5-d_f)}    
\end{equation}\label{eq:muthukumar}
where $G' \sim \omega^\beta$ and  $0 < \beta < 1$, which correspond to   $1.25  < d_f < 2.5$.  The above relation was originally devised for gels at the percolation threshold but has also been shown to be broadly applicable at later timescales,\cite{Aime2018} as used here. A simple least squares fit to figure \ref{FIG1}(c) reveals a weak power law ($\beta$ =0.02).  We also independently verify the value of $\beta$ using the Kramers-Kronig relation: $G'' / G' = tan( \beta \times \pi/2 )$ which is a consequence of the interdependence between G'' and G'.\cite{chambon1987linear} We find $\beta = 0.022\pm0.002$ (figure \ref{SUPP-kramerskronig}, supplementary), in excellent agreement with the power law fit. Taking $\beta = 0.02$, equation (1) indicates a fractal dimension of 2.48. In previous work we have carried out structural characterisation of $I27_5$ hydrogels\cite{da2017assessing}, which revealed dense nanoscale clusters of polyproteins with a mean diameter of 20  nm, surrounded by sparser inter-cluster regions, which may be either folded\cite{Hughes2020} or unfolded\cite{Hughes2021}, depending on the constituent protein. These nanoscale clusters have been shown to be fractal in nature with values of $d_f$ across a broad range 2.1-2.7.\cite{Hughes2020} How then should we compare the value of $d_f$ of the nanoscale clusters to those extracted from time-dependent rheology data?  The rheological response of folded protein hydrogels is likely\cite{AufderhorstRoberts2020} to depend not on the structure of the nanoscale clusters but rather on the connectivity of inter-cluster regions. This is analogous to colloidal gels, in which the hydrogel elasticity is a property that emerges from the density of connections \textit{between} colloidal clusters.\cite{whitaker2019colloidal} Our analysis is therefore reminiscent of other soft matter systems and intriguingly points towards the possibility of a fractal structure that is \textit{length-scale dependent}. Although we cannot infer the precise length scale of the inter-cluster regions through rheology, recent work from our group has measured this length scale for MBP hydrogels as being approximately 100 nm.\cite{Hughes2020} 

\subsubsection*{Hydrogel Structure is Responsive to Deformation}

So far, we have identified that the relaxation spectrum of an $I27_5$ hydrogels can be functionally fitted to a power law model, with exponent $\beta$. To validate this, we independently carry out a series of stress relaxation experiments, in which a step strain is applied to the sample and the sample's shear stress decays over time.  This approach avoids the changes in sample strain that are inherent to creep measurements while also probing a longer range of timescales than oscillatory frequency sweeps. We probe the extent of the  linear viscoelastic regime through a shear strain ramp (figure \ref{fig:LVEtest}), which confirms that the linear viscoelastic regime corresponds to $\gamma$ $<$ 0.1, with only a slightly nonlinear response at higher strains ($\gamma$ $<$ 0.3).  
\begin{figure}[ht]\centering 
	\includegraphics[width=0.5\linewidth]{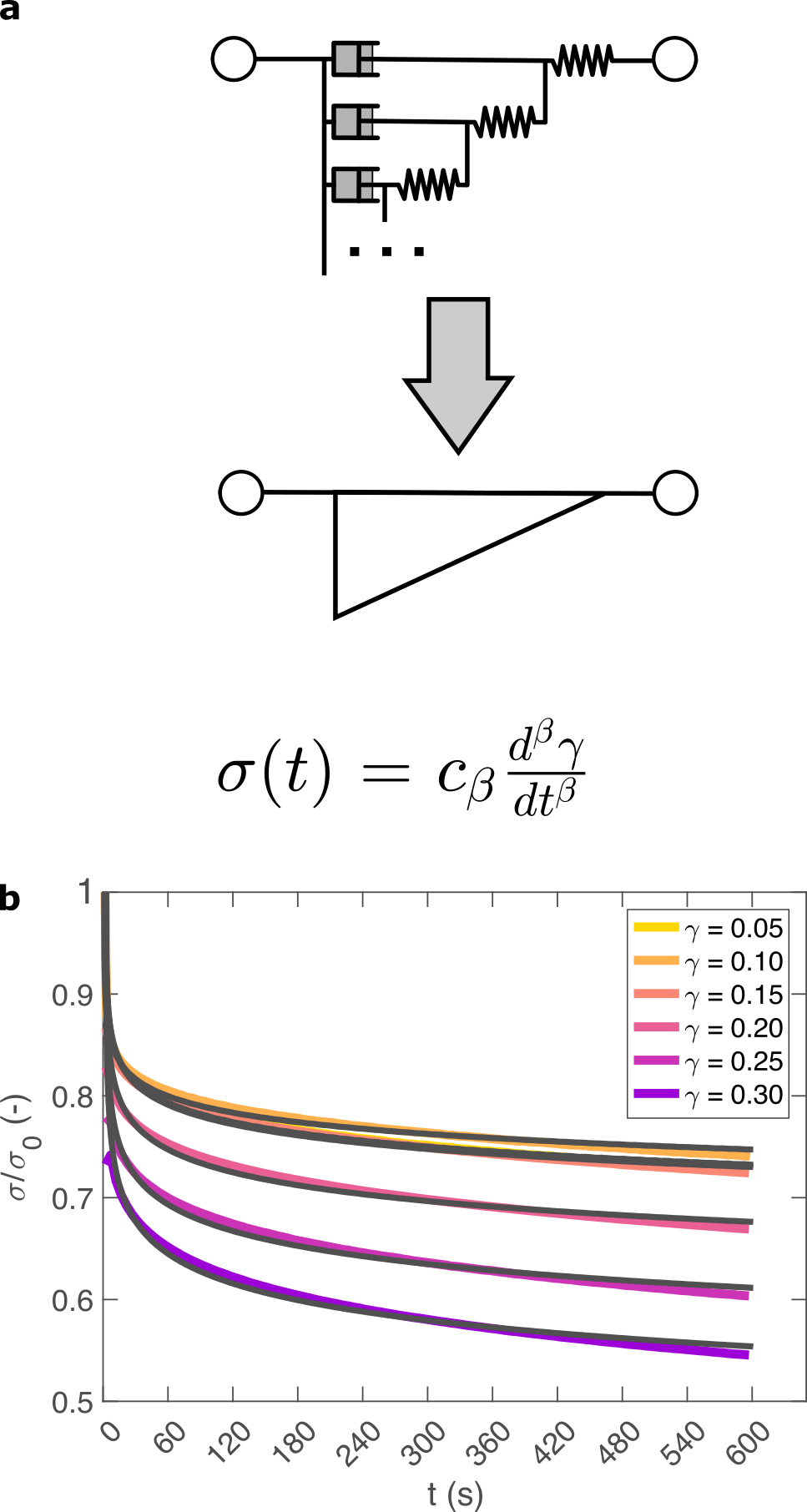}
	\caption{(a) Relaxation data are fitted using a fractional model, which represents an arbitrarily large network of springs and dashpots, using just two parameters $c_\beta$ and $\beta$.  (b) A step strain $\gamma$ at time 0 is applied to an $I27_5$ hydrogel and the normalised stress response  $\sigma$/$\sigma_0$ is probed. The model (grey line) provides an excellent fit across a range of values of $\gamma$. }
	\label{fig:FIG2}
\end{figure}
While the exponent $\beta$ is easily extracted from the frequency sweeps, extracting it from relaxation data requires a mechanical constitutive modelling approach.  Here,  elements of Hookean elasticity (springs) and Newtonian viscosity (dashpots) are combined in an empirical manner to approximate the viscoelastic response.  In many materials, a finite number of such elements are sufficient,\cite{winter1986analysis} however for power law viscoelasticity the spectrum of relaxation times is broad and hence the approach requires a large number of fitting parameters making it unsuitable.\cite{blair1947limitations} To address this, we use a so-called fractional rheological model, which captures  power law rheology using a minimal number of fitting parameters.  This is achieved by reducing an arbitrary large number of springs and dashpots (figure \ref{fig:FIG2} (a)) to a single mathematical element  commonly known as a springpot model \cite{Koeller1984} with two fitting parameters  $\beta$ and $c_\beta$. The parameter $c_\beta$, is often referred to as a  \textit{quasi-property} \cite{blair1947limitations}   since its SI units  (Pa.s$^\beta$) contain a power law exponent.  From a practical perspective $c_\beta$ can most simply be conceptualised as referring to the `firmness' or `strength' of the material \cite{Blair1942}. The springpot model describes stress relaxation $\sigma(t)$ following an applied step strain $\gamma$ over a measurement timescale $t$  as:
\begin{equation}
\sigma(t) = c_{\beta}\frac{d^\beta \gamma}{dt^\beta}
\label{springpot model}
\end{equation}
Here $\beta$ has the same physical meaning as previously defined. The elegance of the above relation is reflected in the fact that it reduces to the equation for a Hookean spring when $\beta = 0$ and to a Newtonian dashpot when $\beta = 1$. The underlying physical basis for this model is supported by the fact that when fractal soft materials are embedded with passive particles, the particle motion is sub-diffusive and anomalous ($<\Delta x^2> \sim t^{\beta}, \beta < 1$).\cite{Sokolov2002, Metzler2000} This sub-diffusive behaviour arises from the multi-scale structure of the material which also directly leads to power law stress relaxation. \cite{muthukumar1985dynamics} As such, the fractional approach has come to be established as the most predictive model for power-law viscoelastic materials.\cite{Koeller1984,Faber2017,schiessel1995generalized,Aime2018} 

\begin{figure}[ht]\centering 
	\includegraphics[width=0.5\linewidth]{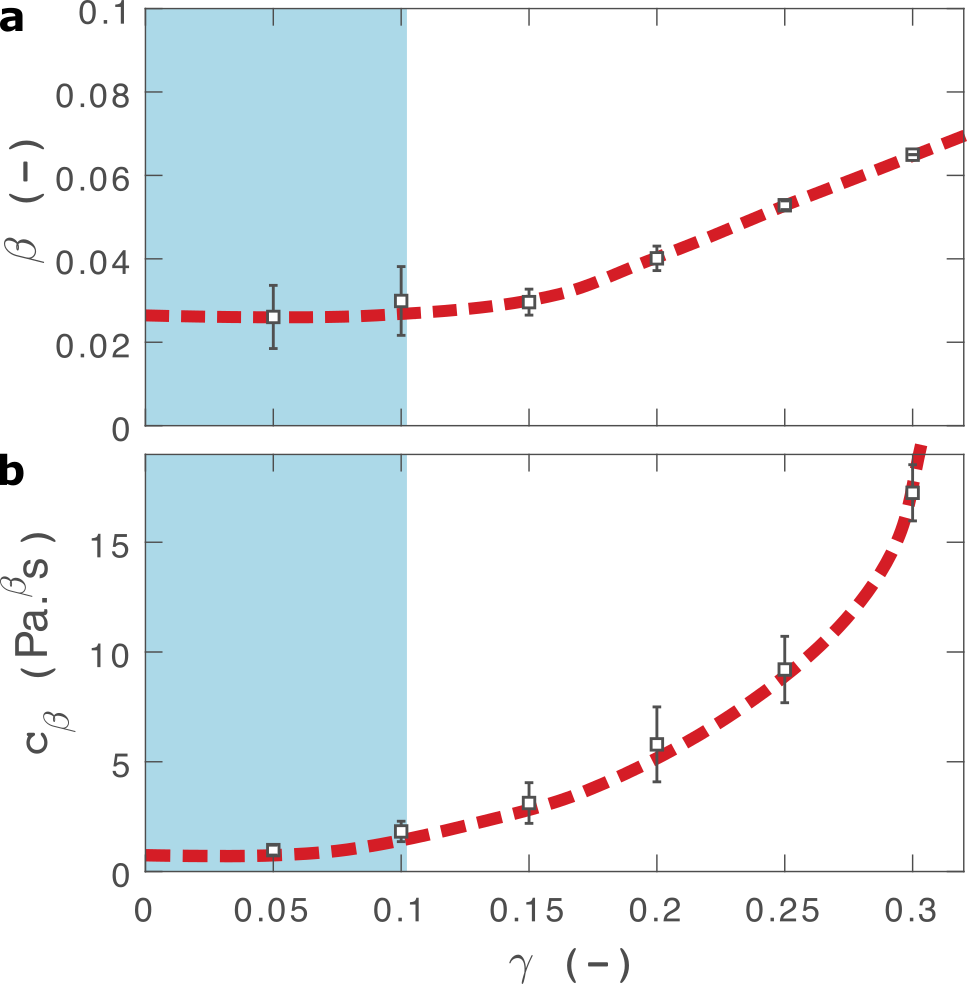}
	\caption{Fractional fit parameters extracted from applying equation \ref{springpot model} to data in shown in figure \ref{fig:FIG2}. The shaded region indicates the linear viscoelastic regime. 	Dashed lines are guides for the eye. Error bars represent fits over multiple measurements}
	\label{fig:FIG3}
\end{figure}

As shown in figure \ref{fig:FIG2}(b), equation \ref{springpot model} provides an excellent fit to the stress relaxation data, both within the linear viscoelastic regime ($\gamma$ $<$ 0.1) and even for the nonlinear regime (0.1 $<\gamma<$0.3 ).  For step strains that exceed this strain (figure \ref{fig:SUPP-relaxationfracture}) we consistently observe an instantaneous decrease in stress during the measurement, suggesting that the sample undergoes irreversible fracture. The fitted parameters $\beta$ and $c_\beta$ are shown in figure \ref{fig:FIG3}.

The fractional model provides an excellent fit to the relaxation data. We also attempted to fit several different variations of the two-component fractional model, incorporating different combinations of springs, dashpots and springpots arranged in series and in parallel in attempt to improve our model. The use of either fractional Maxwell model (two springpots in series, figures \ref{fig:SUPP-fractionalmaxwell}\cite{Holder2018}) or a modified power law model (springpot combined with a dashpot and spring, figure \ref{fig:SUPP-fractionalmodified} \cite{Bonfanti2020}) provide a poor fit to the data as both predict a plateau in $\sigma(t)$ at long timescales. A fractional Kelvin Voigt model (two springpots in parallel, figure \ref{fig:SUPP-fractionalKV}) does provide a close fit to the data, however the additional parameters are degenerate to fitting. 

Within the linear viscoelastic regime, the value of $\beta$ is independent of applied strain, as expected, and has an average measured value of 0.03 $\pm$ 0.01. This is in excellent agreement with the frequency spectra ($\beta=0.02$), demonstrating that the power-law model accurately predicts the viscoelastic material properties of the $I27_5$ hydrogels,   which are likely to originate from a cross length-scale fractal structure.   Interestingly, the springpot model appears to provide an excellent fit to the relaxation data at strains that exceed the linear viscoelastic regime. The reasons for this are not immediately clear since the fractional model that we employ is only strictly valid for linear viscoelastic materials, \cite{surguladze2002certain} although attempts have been made to extend it to nonlinear behaviour.\cite{jaishankar2014fractional,Nicolle2010} It may be that the deviations from linear viscoelasticity for $0.1 < \gamma < 0.3$ (figure \ref{fig:LVEtest}) are sufficiently minor so as not to invalidate the model. We find that $c_\beta$ increases with strain,  which suggests a strengthening of the hydrogel network.  This, accompanied by an increase in $\beta$ over the same range of applied strain, suggests progressive changes in network structure as the strain is increased.  $\beta$ reaches a maximum of $0.065\pm10^{-5}$ for $\gamma$, which using the equation (1), would correspond to a slight decrease in fractal dimension from $d_f = 2.47$ to $d_f = 2.44$. Assuming that our model is indeed valid at higher strains, we propose that this change may reflect the progressive unfolding of the constituent $I27$ domains as the applied strain is increased, leading to  changes in the hydrogel network structure. At the same time, the increase in $c_\beta$ would imply a strengthening or reinforcement of the network at increased strain. One explanation for this is that the polyprotein chains stiffen, by aligning with the direction of shear (leading to an increase in $c_\beta$) while simultaneously unfolding (leading to an increase in $\beta$).

 Comparison can be made to a previous study by our group on crosslinked BSA hydrogels. In that study, the addition of a reducing agent resulted in breakage of covalent bonds within the BSA protein, resulting in a more force labile protein.\cite{Hughes2021} Upon cross-linking, \textit{in situ} protein unfolding was found to define the network architecture and mechanics, with regions of clusters of folded protein interconnected by unfolded protein. The study also noted an increase in the power law exponent when the disulphide bonds that prevent unfolding of the protein were broken through chemical reduction. Remarkably, the power law exponents before and after covalent bond breaking ($0.027 \pm 0.002$ and $0.061 \pm 0.001$, respectively) are almost identical to the exponents at low and high strains measured here ($0.03 \pm 0.01$ and $0.065\pm10^{-5}$, respectively), providing further support for our hypothesis that the structural changes implied by our data arise from the force induced unfolding of individual protein domains. 

\subsubsection*{Nonlinear Mechanics Indicate Reversible Alignment and Unfolding of Polyproteins}

\begin{figure*}[ht]\centering
	\includegraphics[width=1\linewidth]{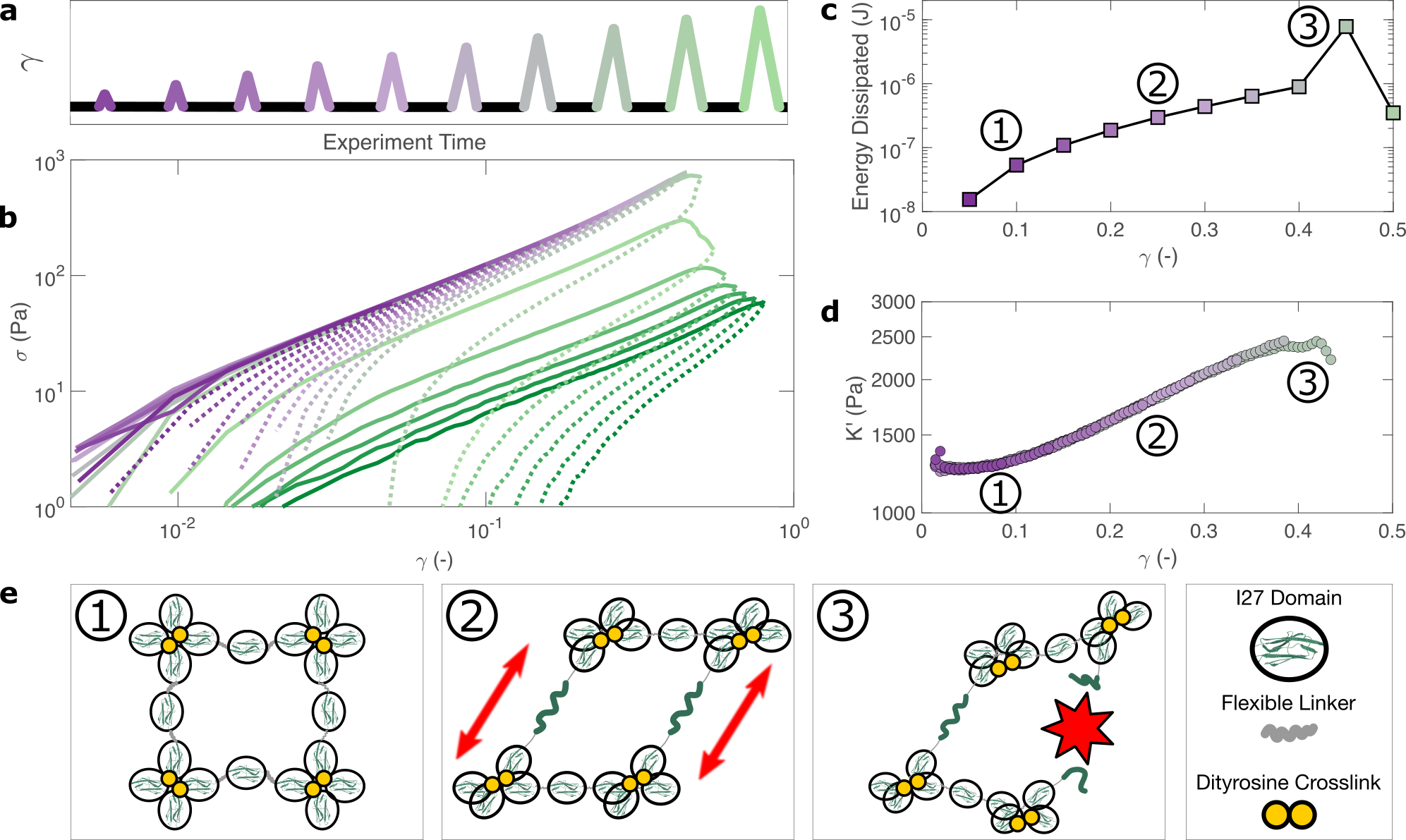}
	\caption{The bulk response of polyprotein networks under large deformation reveals insights into the mechanisms of energy dissipation. A series of triangular strain ($\gamma$) pulses of increasing magnitude are applied (a). For $\gamma < 0.4$ each stress ($\sigma$) curve overlays closely with its predecessor (b) indicating that the hydrogel's mechanical response is fully reversible. Increasing applied strain correlates closely with energy dissipation (c) and strain stiffening (d), which are macroscopic manifestations of the unfolding of individual protein domains and the stretching of polyproteins, respectively. This is shown diagrammatically (e).  At $\gamma = 0.4$ the energy dissipated increases sharply and subsequent loading curves no longer overlay, indicating the irreversible  fracture of the hydrogel.}
	\label{fig:FIG4}
\end{figure*}

How do the structural properties of $I27_5$ hydrogels that we have encountered, change under deformation? Our relaxation data show intriguing clues that suggest network strengthening and unfolding under applied strain.  To probe this independently we carry out nonlinear cyclic deformation experiments\cite{Schmoller2010,Kurniawan2016}. As depicted in figure \ref{fig:FIG4}(a) and figure \ref{fig:SUPP-repeatedstressstrain}, the $I27_5$ polyprotein network is loaded and unloaded up to a threshold strain $\gamma_{max}$ through a triangular strain pulse at a constant shear rate and the magnitude of $\gamma_{max}$ is increased with each subsequent shear pulse. 

Within the linear viscoelastic regime, ($\gamma_{max} < 0.1$) the loading and unloading curves (figure \ref{fig:FIG4} (a), solid and dashed lines respectively) appear to overlay closely suggesting a reversible and purely elastic response. However as $\gamma_{max}$ increases, a deviation between the loading and unloading curves begins to emerge.  Calculating the energy loss in each cycle (\ref{fig:FIG4}(c)) from the ratio of the areas beneath the loading and unloading curves (see methods), shows that inelastic energy dissipated during loading steadily increases with $\gamma_{max}$. Furthermore, each subsequent loading curve appears to  overlay with its predecessor, indicating that this inelastic dissipation does not bring about any permanent changes to the network stiffness. This complete reversibility of the network is unusual as most polymer networks that exhibit reversibility, such as filled rubbers\cite{Diani2009} and biological tissues also exhibit a well-defined residual cyclic softening. By contrast, we observe an almost perfect overlay between loading curves  for strains up to $\gamma = 0.4$.
The profile of the loading curves is also revealing of changes in hydrogel stiffness during deformation. For $\gamma > 0.1$, the stress $\sigma$ required to increase the strain becomes progressively higher, indicating strain-stiffening.\cite{Storm2005} This can be most clearly seen in the evolution of the differential elastic modulus $K'$ with respect to strain (figure \ref{fig:FIG4}(d)), which increases  approximately by a factor of 2 before the onset of rupture. As with the loading curves, the relationship between $K'$ and $\gamma$ overlays closely across all curves, confirming that no softening takes place between loading cycles. Plotting $K'$ with respect to $\sigma$, reveals that the stiffening follows a power law relationship of $K'\sim \sigma^{0.6}$ (figure \ref{fig:SUPP-Kvssigma}). Such power law stiffening is a general feature of semiflexible networks for which the distance between crosslink nodes is similar to the polymer persistence length and arises from the entropic resistance of thermal undulations against stretching.\cite{Storm2005}  To assess whether our system can be considered to be semiflexible we carry out an analysis of the polyprotein dimensions (see supplementary information) which reveals that the distance between the crosslink sites is $5.44$  nm. To calculate the persistence length of the polyproteins we draw on our previous computational calculations for a range of polyprotein dimensions. \cite{Hanson2019} The most important parameter here is the ratio between the length of the linker domain and the radius of the globular domains, which for our dimensional analysis is $\approx 0.34$. This corresponds\cite{Hanson2019} to a persistence length in the range of $\approx$ 7-14  nm. Therefore, the persistence length and crosslink length are of the same order of magnitude and the polyprotein can be considered as semiflexible. For semiflexible networks it is expected that $K'\sim \sigma^{1.5}$\cite{MacKintosh1995} although exponents as low as $0.5$\cite{Tharmann2007} have been observed elsewhere and are thought to arise from inelastic effects such as crosslink unbinding or axial stretch. 

For our system, strain stiffening appears to occur concurrently with energy dissipation, which agrees with the stiffening and dissipation hypothesis that emerges from our relaxation data. From the strain ramp data, it seems likely that strain stiffening occurs due to the stretching of the constituent semiflexible polyproteins and that these polyproteins undergo force-induced unfolding concurrently (figure \ref{fig:FIG4} (e)). 

\subsubsection*{Fracture of Polyprotein Hydrogels is Driven by Structural Adaption}

If the unfolding and refolding of $I27_5$ in the hydrogel is indeed reversible then how does the fracture, as observed from strains exceeding 0.4, take place? To investigate  this further, we carry out oscillatory frequency sweep measurements after each of the  triangular strain pulse measurements. Each frequency sweep therefore represents the linear viscoelastic properties of the hydrogel after each cycle of deformation. 
\begin{figure}[ht!]\centering 
	\includegraphics[width=0.5\linewidth]{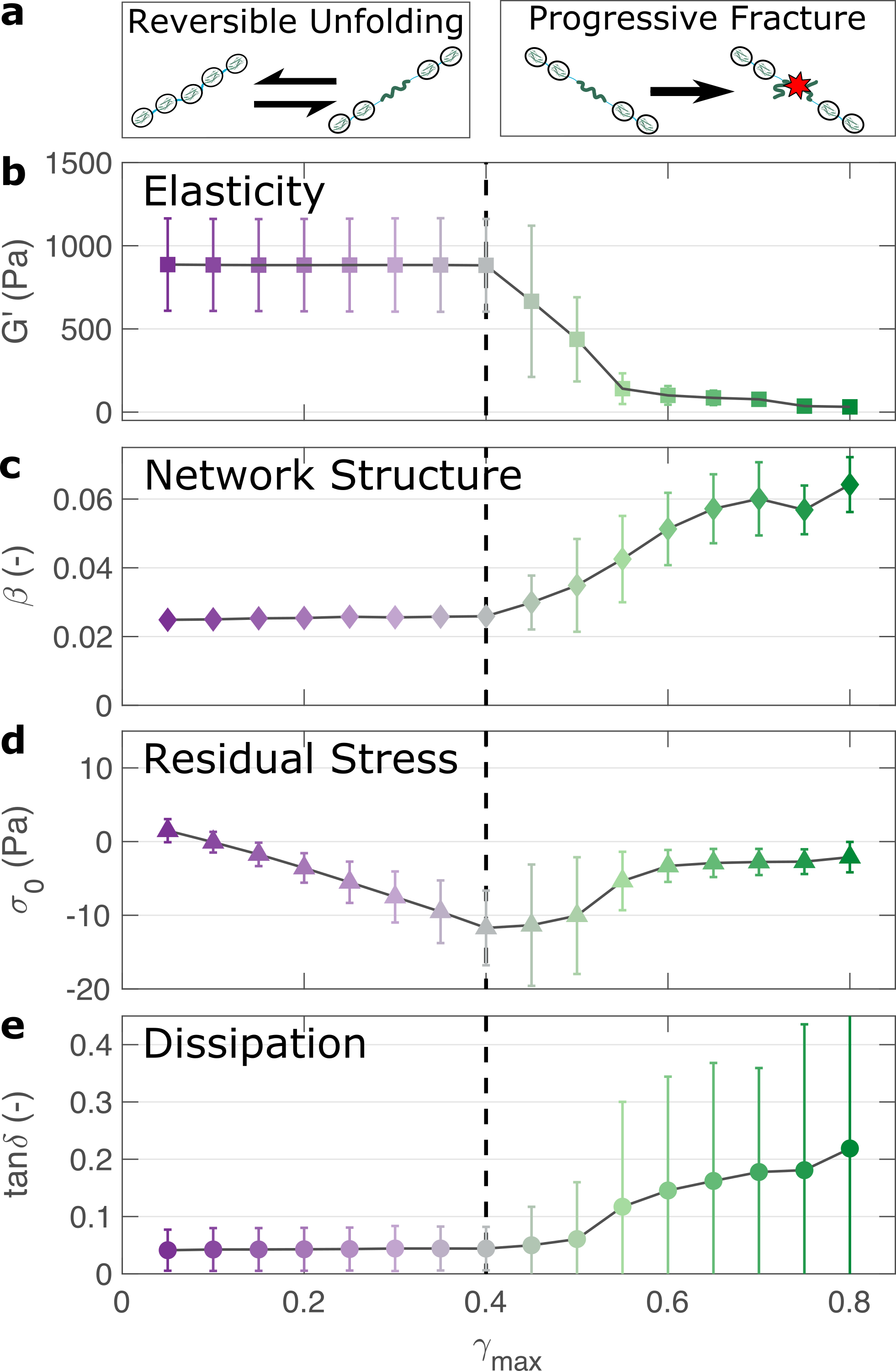}
	\caption{The linear viscoelastic properties of the $I27_5$ hydrogels after strain pulse deformation cycles of magnitude $\gamma_{max}$ reveal the fracture properties of the hydrogel network (a). The elastic modulus remains unchanged below a threshold strain of $\gamma_{max} = 0.4$, after which it steadily decreases, indicating network fracture (b). The power law exponent $\beta$, which represents the fractal structure of the hydrogel increases after the fracture point, indicating structural changes post-fracture (c). The negative stress $\sigma_{0}$ needed to return the network to zero strain after each deformation cycle steadily increases before fracture, suggesting a build-up of tension after each deformation cycle (d). Meanwhile the network dissipation (tan$\delta$ = G''/G') remains relatively low over all values of $\gamma_{max}$, indicating that bond rupture rather than protein unfolding is the underlying cause of the fracture of the hydrogel network above a threshold strain of $\gamma_{max}$=0.4.}
	\label{fig:structureadaption}
\end{figure}
The profiles of the individual frequency sweeps of the hydrogels remain highly reproducible at pre-fracture strains, (figure \ref{Supp - postloadingfreqsweeps}) which is consistent with our previous observation that the deformations are fully reversible. At higher strains, the linear elastic modulus $G'$ begins to decrease (figure \ref{fig:structureadaption} (a)). Over the same strain range, the power law exponent of the frequency sweeps $\beta$ increases,(figure \ref{fig:structureadaption} (b)) indicating permanent structural change, even after the removal of strain.  At this final stage, the value $\beta$ reaches  $0.064\pm0.008$. Remarkably, this is identical to the maximum value of $\beta$ measured through stress relaxation experiments (Figure \ref{fig:FIG3}). 

The fracture of biopolymer networks is a complex phenomenon that is challenging to model.\cite{Jung2015} In network-forming systems, fracture typically occurs suddenly and is preceded by a dynamic microscopic changes\cite{Cipelletti2020}. The eventual mechanism of fracture can originate from the dissociation of noncovalent bonds, the permanent rupture of covalent bonds, or both concurrently.\cite{Ackbarow2007} We expect that covalent bond rupture is the most likely cause in our system since the hydrogels are covalently crosslinked. Furthermore, the high energy dissipation at $\gamma_{max} < 0.4$ (figure \ref{fig:FIG4}) implies protein unfolding occurs pre-rupture. This observation is consistent with the unfolding forces of an $I27$ domain ($\sim$ 0.2 nN\cite{best2003mechanical}), which are lower than the typical rupture forces of a covalent bond ($\sim$ 1-2 nN\cite{Grandbois1999} at a pulling speed of 10 nN.s$^{-1}$). One possible mechanism for fracture is that as the polyprotein approaches a fully unfolded state, the tension that was previously directed to domain unfolding instead is applied to the unfolded polypeptide chain leading ultimately to irreversible peptide bond rupture. It is intuitive that such rupture would occur at the weakest bond in the chain, in this case a peptide bond in either the  unfolded $I27$ domains or the linker region.  Further evidence for the mechanism of rupture can be found by re-examining the strain pulse data in figure \ref{fig:FIG4}. Specifically, the stress required to return the hydrogel to zero strain after each pulse is denoted $\sigma_0$ (shown diagrammatically in figure \ref{fig:SUPP-remodellingparam}). This represents the residual stress after each deformation cycle. $\sigma_0$ is initially close to 0 and becomes more negative after each cycle until the onset of fracture (figure \ref{fig:structureadaption}(c)). Post fracture, $\sigma_0$ no longer becomes more negative, which may reflect the loss of tension in the network as the polyprotein chains rupture. Finally, the value of tan$\delta$ (figure \ref{fig:structureadaption} (d)), which quantifies the relative viscous dissipation of the network in the form of the ratio between the elastic and viscous moduli, remains relatively low ($\approx 0.2$, at the highest values of $\gamma_{max}$). Given that the chains are highly dissipative at high strains (figure \ref{fig:FIG4} (d)) and given the theoretical range of tan$\delta$ (0-1), a much higher dissipation would be expected if the polyprotein chains remained intact but unfolded. It is therefore likely that the network fracture is determined by the reversible unfolding of the protein domains which, at significantly high tension leads to eventual bond rupture as shown diagrammatically in figure \ref{fig:structureadaption} (e). 

\subsection*{Discussion}
\addcontentsline{toc}{section}{Discussion and Summary} 
In summary, we have  applied a multi-modal rheology approach to interrogate the viscoelastic properties of folded protein hydrogels.  Using these approaches, we highlight several important and previously unappreciated aspects of their material properties.  

Firstly, we determine that the viscoelasticity of $I27_5$ hydrogels can be accurately modelled by a simple two-parameter fractional viscoelastic constitutive model. This model manifests itself as a power law in both the frequency response (figures \ref{FIG1} (c) and \ref{Supp - postloadingfreqsweeps}) and in the time-dependent stress relaxation response (figure \ref{fig:FIG2}). From a practical perspective, this indicates that hydrogel's stress relaxation is governed not by a single characteristic timescale but rather by a broad spectrum of timescales. This is perhaps surprising as it suggests that the viscoelastic response of the hydrogel, at least within the linear viscoelastic regime, is not governed primarily by the  narrow and well-defined unfolding timescale of the $I27$ protein domains when stretched axially.\cite{Rief1997}  Instead the architecture of the hydrogel appears to result in a more complex spectrum of relaxation timescales, which likely arises from the orientation and arrangements of the protein domains. There are several possible mechanisms to explain this behaviour. Firstly, it could be argued that the formation of the hydrogel network causes partial unfolding, however we show a reduction of the fraction of folded protein of just 12$\%$ (figure \ref{FIG1}). A more likely explanation is that the fractal structure of the hydrogel\cite{Hughes2020,Hughes2021,AufderhorstRoberts2020} leads to a broad spectrum of relaxation modes, which reflects the broad spectrum of length scales within the network.\cite{Metzler2002} One additional contributing factor could be the random alignment of the polyprotein chains within the hydrogel network in relation to the direction of shear deformation. In principle this could lead to changes in the relaxation timescales as the mechanical stiffness of folded proteins is known to depend on their pulling geometry\cite{Dietz2006}. The broad spectrum of relaxation times we observe here appears to be common in polymeric systems\cite{winter1997}, cells\cite{Fabry2001,Bonfanti2020} and tissues\cite{kohandel2005}. Here, we show that this behaviour also extends to folded protein hydrogels.  

Using a fractional viscoelastic model, we show that the power law exponent $\beta$ is remarkably consistent between the oscillatory shear and stress relaxation experiments used in this study.  Therefore, we may say that power law viscoelasticity is not an artifact of any particular technique  but is instead an intrinsic property of the hydrogel and is likely to reflect the hydrogel's fractal structure. A consistent value of $\beta$ is also measured after multiple cycles of deformation, indicating that this fractal structure is robust to perturbation. More surprisingly, our model also closely fits the relaxation response at larger deformations in the nonlinear viscoelastic regime (figure \ref{fig:FIG3}(a)). While our model is strictly valid only for linear viscoelasticity, we note that the fitted parameter $\beta$ in this regime is close to the value of $\beta$ as probed by oscillatory shear after the hydrogel fracture. We speculate that this may be indicative of permanent changes to the hydrogel's structure at large deformations.  Modelling this behaviour using more advanced constitutive modelling that accounts directly for nonlinear effects\cite{jaishankar2014fractional,Nicolle2010} would an interesting topic of further study. 

Within this nonlinear regime we observe that the hydrogel undergoes stiffening while subjected to shear strains, which is manifested both in relaxation data (figure \ref{fig:FIG3} (b)) and in strain ramp data (figure \ref{fig:FIG4} (d)). In line with predictions,\cite{Hanson2019} this is consistent with the semiflexible nature of the polyprotein chain, a property that arises directly from the dimensions of the folded proteins. Recreating strain stiffening in synthetic hydrogels is a growing area of research\cite{Jaspers2014} and although the degree of stiffening is relatively low for this system (a twofold increase in modulus before fracture) it is likely that a modification of the polyprotein dimensions could tune and enhance this behaviour further. 

This strain-induced stiffening occurs at low strains, which corresponds to the onset of nonlinearity in stress relaxation experiments (figure \ref{fig:FIG2}) and to an observed increase in energy dissipation (figure \ref{fig:FIG4} (d)). Our relaxation data indicates an increase in  the power law exponent  $\beta$ during deformation but notably there is no permanent changes to the hydrogel’s viscoelasticity or $\beta$ (figure \ref{fig:structureadaption}), suggesting that the network structure is not significantly altered by repeated deformation. This stands in direct contrast to conventional rubber materials, in which the viscoelastic response from each subsequent deformation cycle is directly determined by the strain history of the sample.\cite{Diani2009} It is likely that this behaviour is a consequence of the $I27$ domain's propensity to re-fold upon removal of stress, since the same behaviour has been observed for hydrogels constructed from the similarly reversible GB1 domains.\cite{Lv2010}  Nevertheless, our analysis of the repeated loading curves indicates that some stress accumulates in the network after each deformation cycle (figure \ref{fig:structureadaption}(d)), which likely contributes to its eventual fracture. 

Finally, we wish to re-emphasise that the weak power-law rheology that we observe here is also commonly observed in biological tissue.  Indeed, slow internal relaxation processes appear to be intrinsic to a broad range of living materials \cite{Nicolle2010}. The mechanical similarities between biological tissues and folded protein hydrogels are apparent and may provide further support for their potential applications in replacing and repairing living tissue. 

In summary, we have shown that the viscoelastic response of $I27_5$ folded protein hydrogels is governed not only by the molecular mechanics of the folded protein domains, as has commonly been assumed, but also by their meso-scale structure. Within the linear viscoelastic regime, the protein domains remain folded, and the response follows a characteristic power-law rheological behaviour. This is well described by a fractional viscoelastic model and most likely arises from the hydrogel's fractal structure. At higher strains, nonlinear viscoelasticity is observed in the form of strain stiffening, which is accompanied by an increase in energy dissipation as the polyprotein chains align and unfold. This unfolding shows an excellent degree of reversibility, which affirms the ability of the constituent $I27$ domains to return to their folded confirmation upon removal of stress.  Overall, we show that folded protein hydrogels exhibit a rich diversity of viscoelastic properties which reflect the hierarchy of length-scales, from their energy-dissipative properties (at the length scale of individual proteins) to their stiffening (at the length scale of polyprotein chains) and their linear viscoelasticity and fracture properties (at the length scale of the hydrogel network). We expect that these insights will be instructive in the continuing efforts to modify and control the mechanics of folded protein hydrogels as well as for their envisaged applications as biomaterials.



\bibliography{main.bib}

\bibliographystyle{unsrt}

\textbf{Acknowledgments}: We thank David Head, Ben Hanson, and Matt Hughes for helpful discussions. \textbf{Funding:} We gratefully acknowledge funding from the Engineering and Physical Sciences Research Council with grant EP/P02288X/1. The CD spectropolarimeter was funded by the Wellcome Trust with grant WT094232.  \textbf{Author contributions:} A.A-R, L.D and D.J.B conceived the experiments. A.A-R performed the experiments, analysed the data and wrote the manuscript with contributions from all other authors. S.C expressed and purified proteins. L.D. is the principal investigator and corresponding author. \textbf{Competing interests:} The authors declare that they have no competing interests. \textbf{Data and materials availability:} All data needed to evaluate the conclusions in the paper are present in the paper and/or the Supplementary Materials.


\newpage
\onecolumn

\renewcommand{\thesection}{S\arabic{section}}  
\renewcommand{\thetable}{S\arabic{table}}  
\renewcommand{\thefigure}{S\arabic{figure}} 
\setcounter{figure}{0}

\begin{center}
\Large{Supplementary Materials for}\\
\vspace{12pt}
\Large{\textbf{Power Law Rheology of Folded Protein Hydrogels}}\\
\vspace{12pt}
\large{Anders Aufderhorst-Roberts, Sophie Cussons, David J. Brockwell, Lorna Dougan$^{\ast}$}\\
\vspace{12pt}
$^{\ast}$Corresponding author. Email: l.dougan@leeds.ac.uk
  \end{center}

\vspace{36pt}
\large{\textbf{This PDF file includes:}}\\
Supplementary Text\\
Figs. S1 to S12\\
\newpage

\subsection*{Supplementary Text}
\subsubsection*{Analysis of Polyprotein Dimensions}
\begin{figure*}[ht]\centering 
\includegraphics[width=1\linewidth]{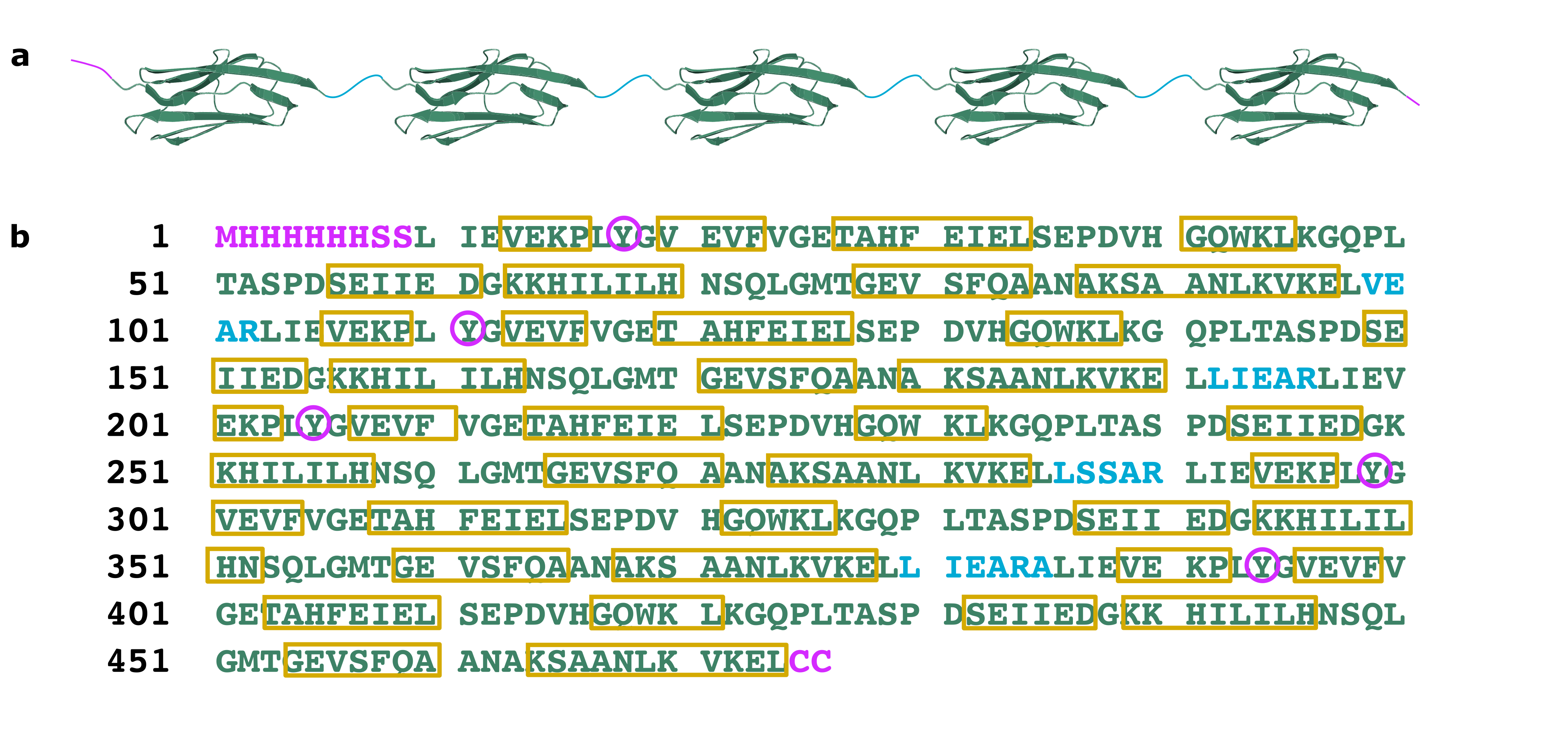}
    \caption{Secondary structure (a) and protein sequence (b) of the $I27_5$ polyprotein used in this study. The molecule comprises five mutated $I27$ domains concatenated with unstructured linker regions (light blue). The secondary structure is predominantly composed of $\beta$-sheets (rectangles) and contains 5 tyrosine residues, (ringed by purple circles) which act as crosslink sites}
    \label{supp-proteindesign}
\end{figure*}

\begin{figure*}[ht]\centering 
\includegraphics[width=1\linewidth]{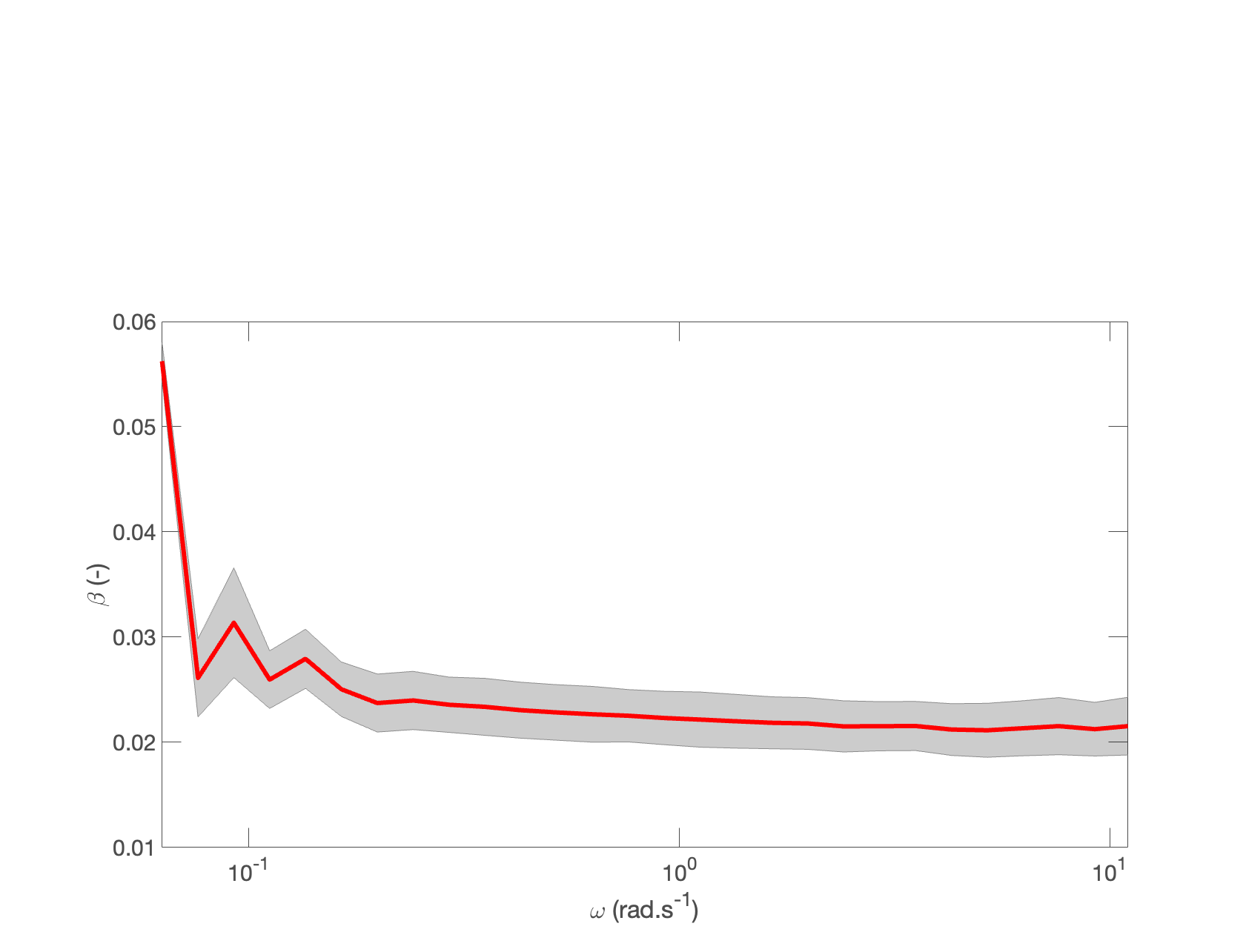}
    \caption{Power law exponent $\beta$ as verified independently using the Kramers-Kronig relation of G'' and G'.  For $\omega>1~\mathrm{rad}~s^{-1}$, $\beta$ is shown to plateau to an average value of $0.022\pm0.002$, in excellent agreement with power laws fitted to $G'$ in figure \ref{FIG1}. Error bars (shaded regions)  represent averaged measurements.}
    \label{SUPP-kramerskronig}
\end{figure*}

\begin{figure*}[ht]\centering 
\includegraphics[width=1\linewidth]{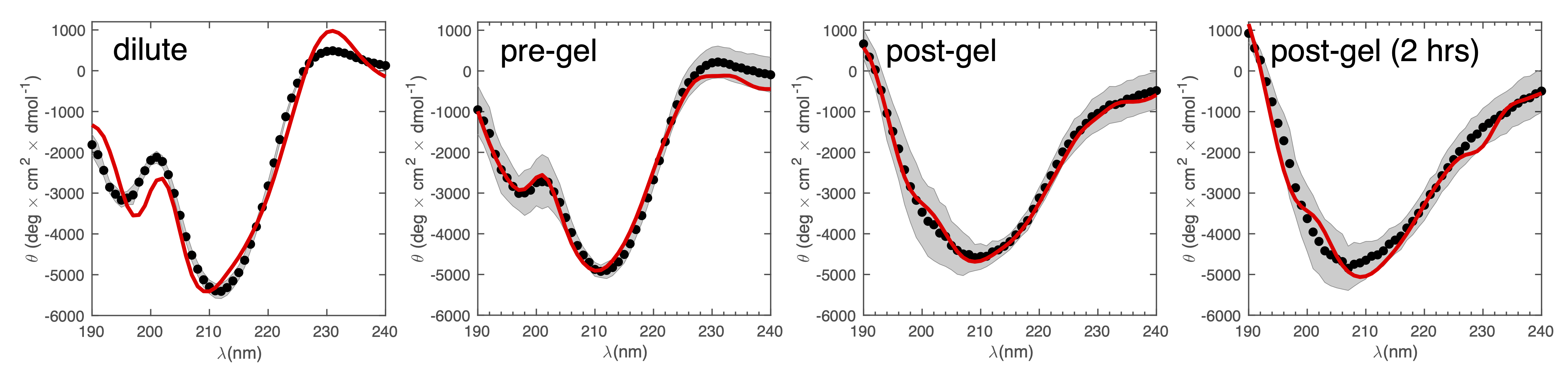}
    \caption{Best fits (solid red lines) to the mean residue ellipticity of $I27_5$ hydrogels (black circles) at dilute concentrations (5 $\mu$M) and at each stage of the gelation process. Fits are carried using the CONTINLL parameter set in Dichroweb.  Error bars (shaded regions)  represent averaged measurements across different samples. }
    \label{SUPP-dichroweb}
\end{figure*}

\begin{figure*}[ht]\centering 4
\includegraphics[width=0.7\linewidth]{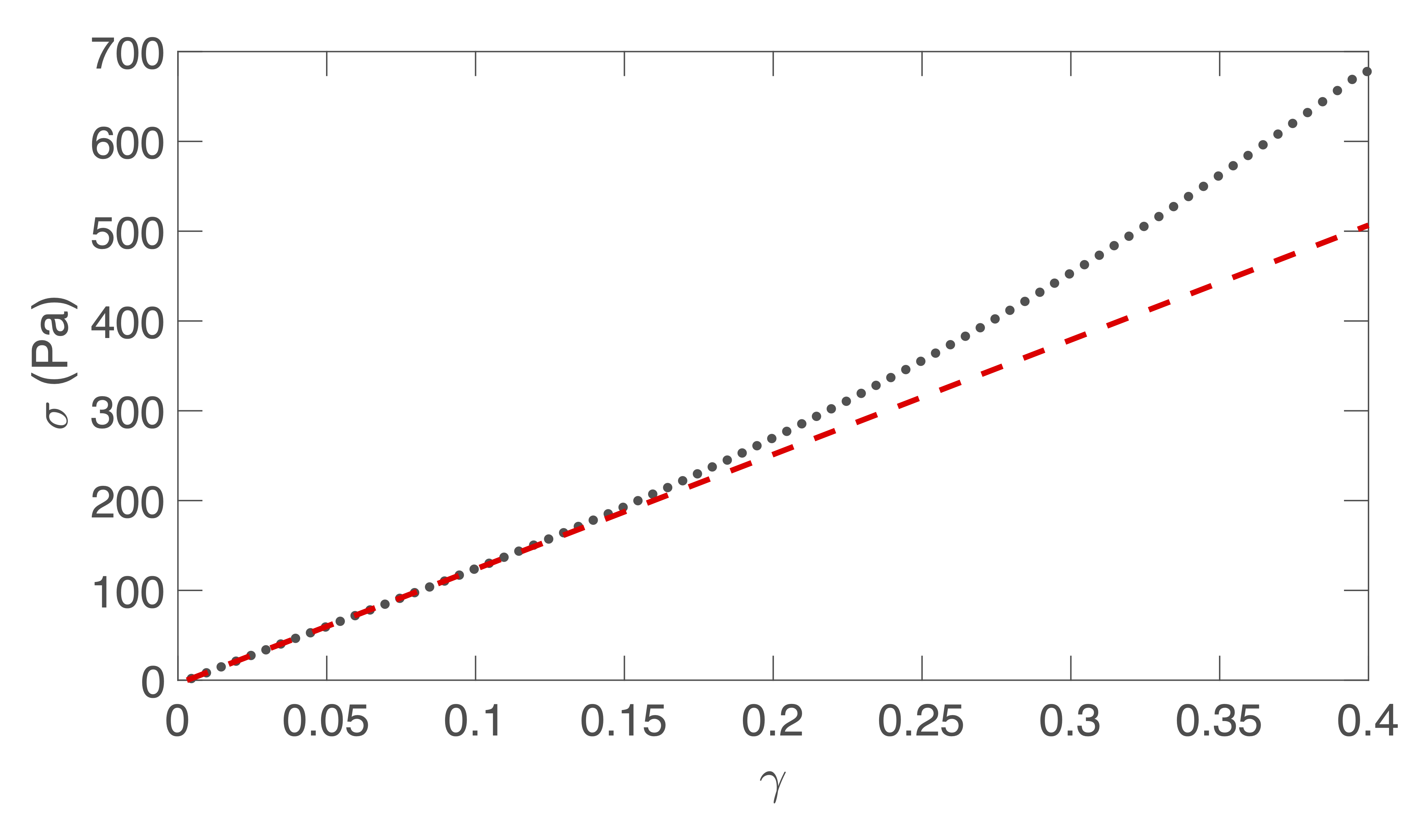}
    \caption{Representative stress-strain Curve of an $I27_5$ hydrogel, captured at a shear strain rate of 0.01 s$^{-1}$ (black dotted line).  Excellent agreement with a linear viscoelastic response (red dashed line) is observed for strain $\gamma < 0.1$. }
    \label{fig:LVEtest}
\end{figure*}

\begin{figure*}[ht]\centering 
\includegraphics[width=1\linewidth]{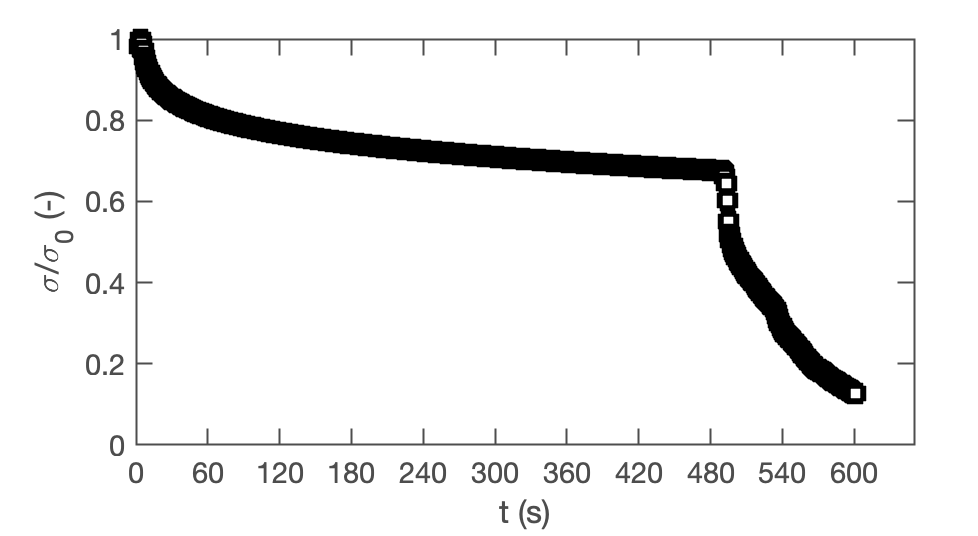}
    \caption{Representative stress relaxation curve at high step strain ($\gamma$ = 0.35). The relaxation curve is discontinuous, indicating that the sample undergoes irreversible fracture. }
    \label{fig:SUPP-relaxationfracture}
\end{figure*}

\begin{figure*}[ht]\centering 
\includegraphics[width=1\linewidth]{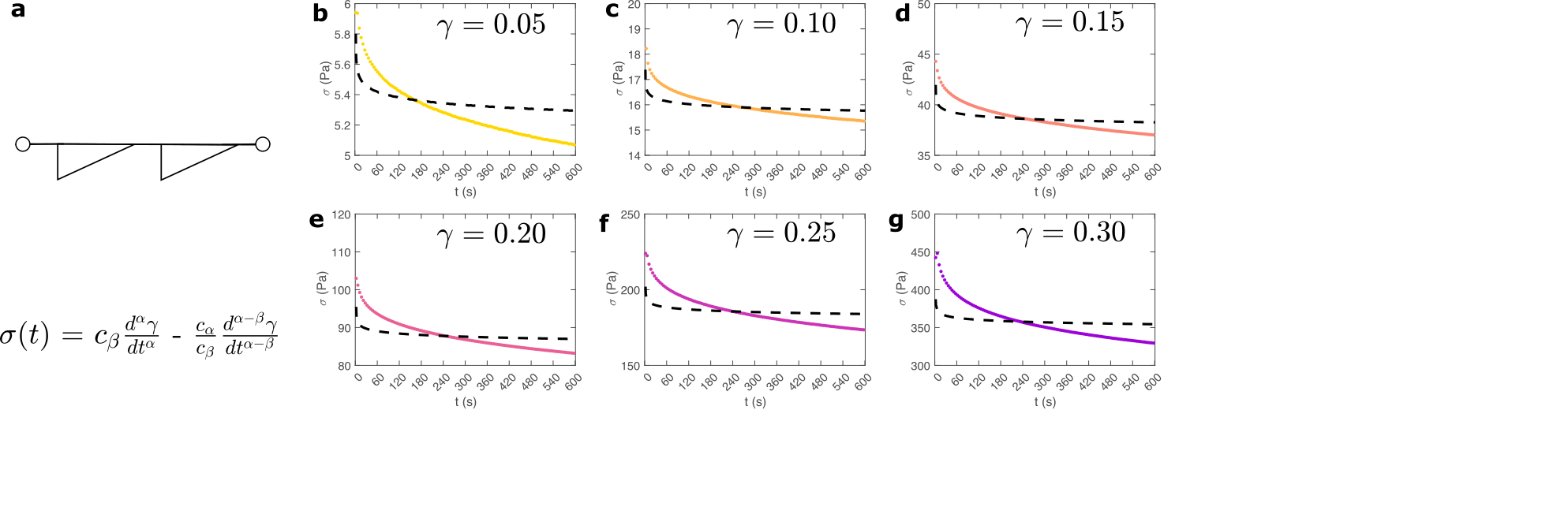}
    \caption{Best fit of a fractional maxwell model (a) to stress relaxation data at a range of different applied strains. Model fits are shown as dashed lines.}
    \label{fig:SUPP-fractionalmaxwell}
\end{figure*}

\begin{figure*}[ht]\centering 
\includegraphics[width=1\linewidth]{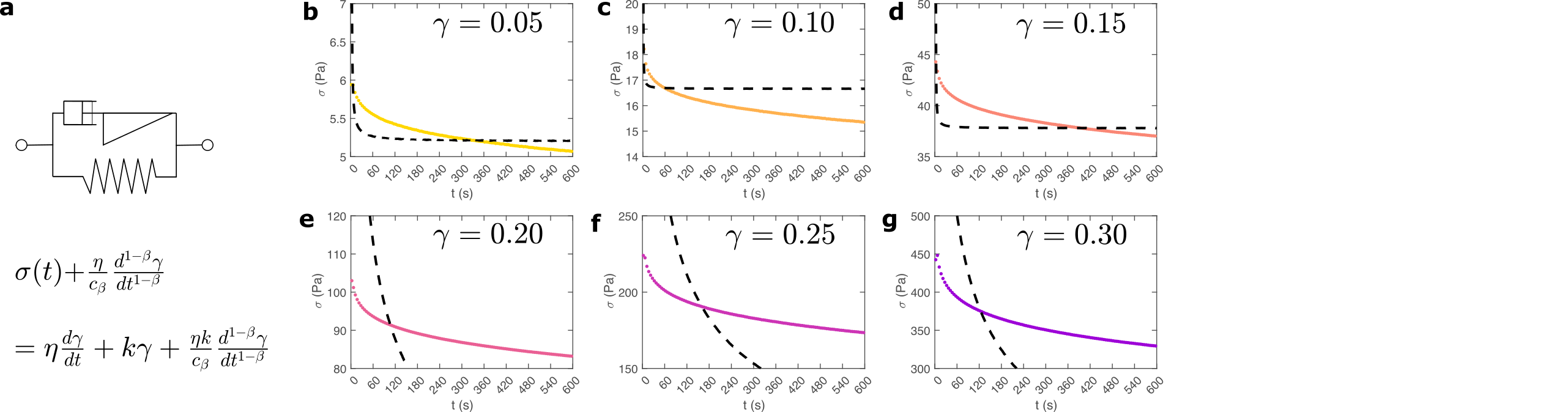}
    \caption{Best fit of a modified Fractional viscoelastic model comprising a fractional springpot element in series with a spring and in parallel with a dashpot (a). The fit is applied to stress relaxation data at a range of different applied strains (b)-(g). Model fits are shown as dashed lines.}
    \label{fig:SUPP-fractionalmodified}
\end{figure*}

\begin{figure*}[ht]\centering 
\includegraphics[width=1\linewidth]{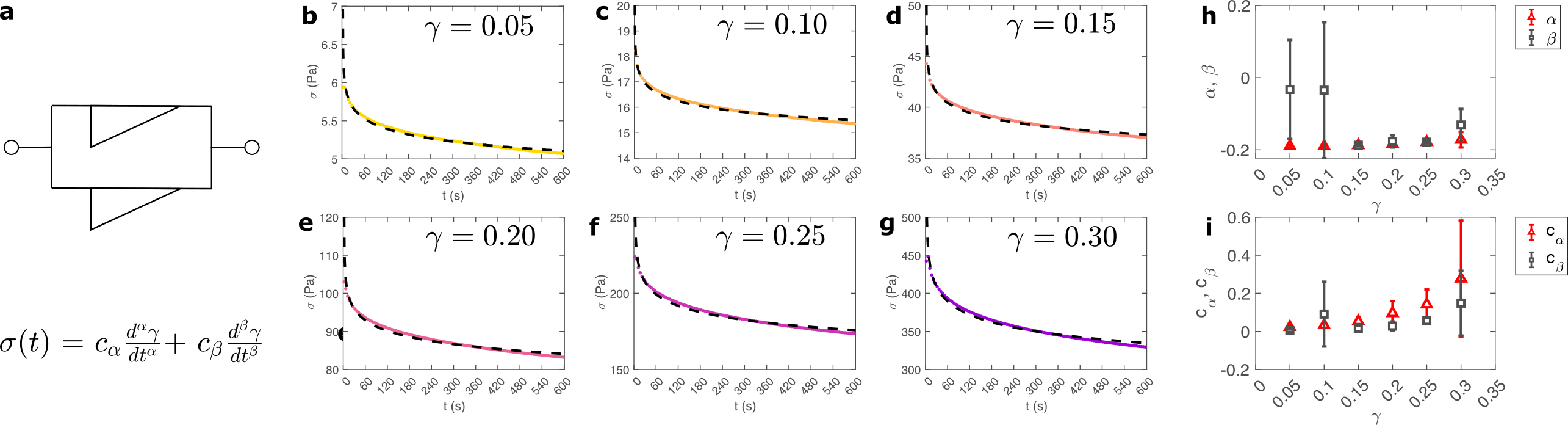}
    \caption{Best fit of a fractional Kelvin Voigt model comprising two springpot elements in parallel.  (a). The fit is applied to stress relaxation data at a range of different applied strains (b)-(g). While the model fits the data closely, the parameters $c_\alpha$ and $\alpha$ are within the error margins of parameters $c_\beta$ and $\beta$ (h)-(i) suggesting that additional parameters cause the fitting to become degenerate.   Error bars represent averaged measurements across different samples.  Model fits are shown as dashed lines.    }
    \label{fig:SUPP-fractionalKV}
\end{figure*}

\begin{figure*}[ht]\centering 
\includegraphics[width=1\linewidth]{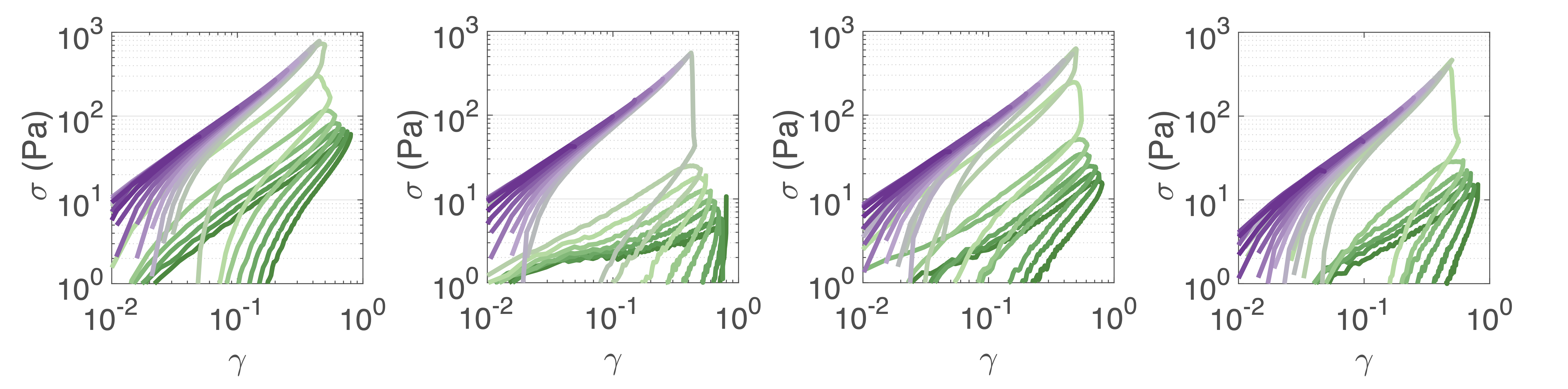}
    \caption{Representative cyclic loading data showing measured stress with respect to applied strain. Solid lines are loading curves, dashed lines are unloading curves. Colours indicate different loading cycles.  Repeat measurements show largely identical nonlinear response, with only slight differences in the network fracture point, denoted by a sudden decrease in stress. }
    \label{fig:SUPP-repeatedstressstrain}
\end{figure*}

\begin{figure*}[ht]\centering 
\includegraphics[width=0.7\linewidth]{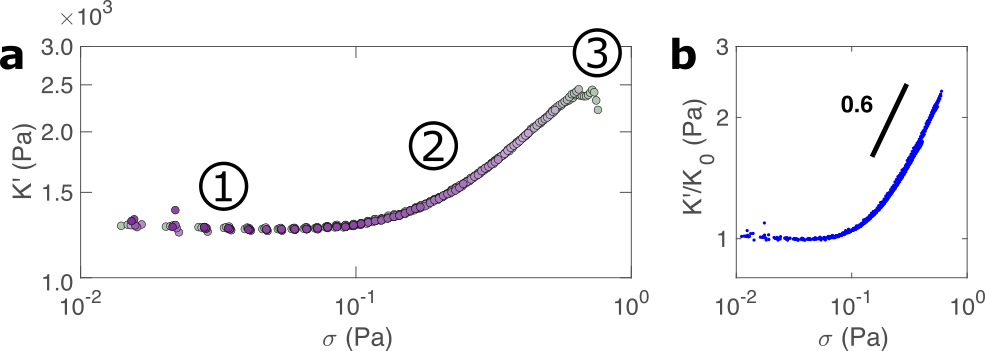}
    \caption{Differential elastic modulus K' with respect to the sample stress, extracted from multiple loading cycles (a). Colours and regimes of mechanical response (1-3) are equivalent to figure \ref{fig:FIG4}.  Once K' is normalised to account for slight differences in the linear modulus $G_0$, data from all repeat measurements superimpose perfectly on to a single curve (b). The hydrogel stiffens as  $K' \sim \sigma^{0.6}$.}
    \label{fig:SUPP-Kvssigma}
\end{figure*}

\begin{figure*}[ht]\centering 
\includegraphics[width=0.7\linewidth]{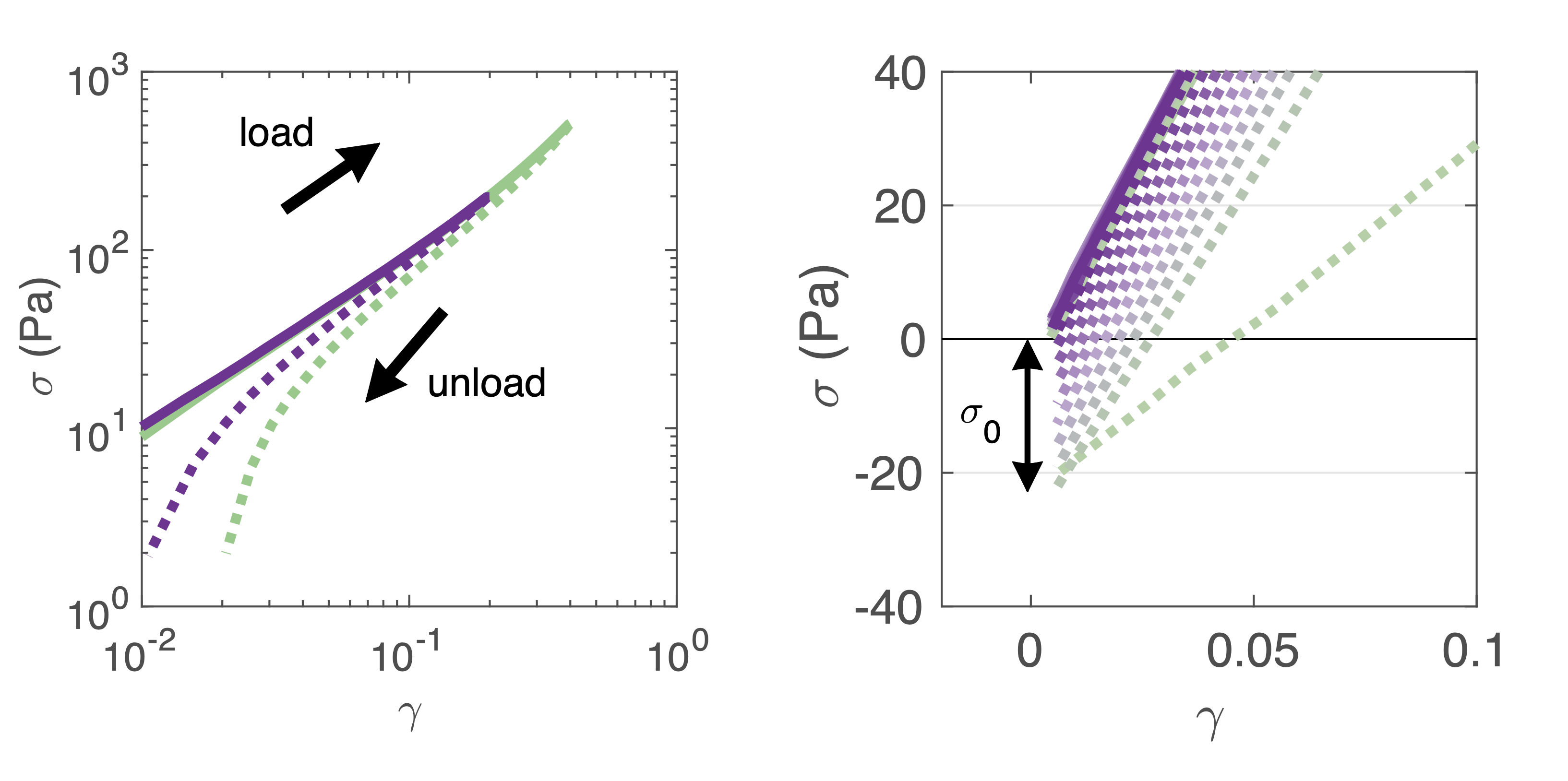}
    \caption{Definition of the remodelling parameter $\sigma_0$. As the network is progressively remodelled after repeated loading cycles, a negative stress $\sigma_0$ required to return the network to zero strain. An increase in $\sigma_0$ indicates an increase in network remodelling. }
    \label{fig:SUPP-remodellingparam}
\end{figure*}

\begin{figure*}[ht]\centering 
\includegraphics[width=1\linewidth]{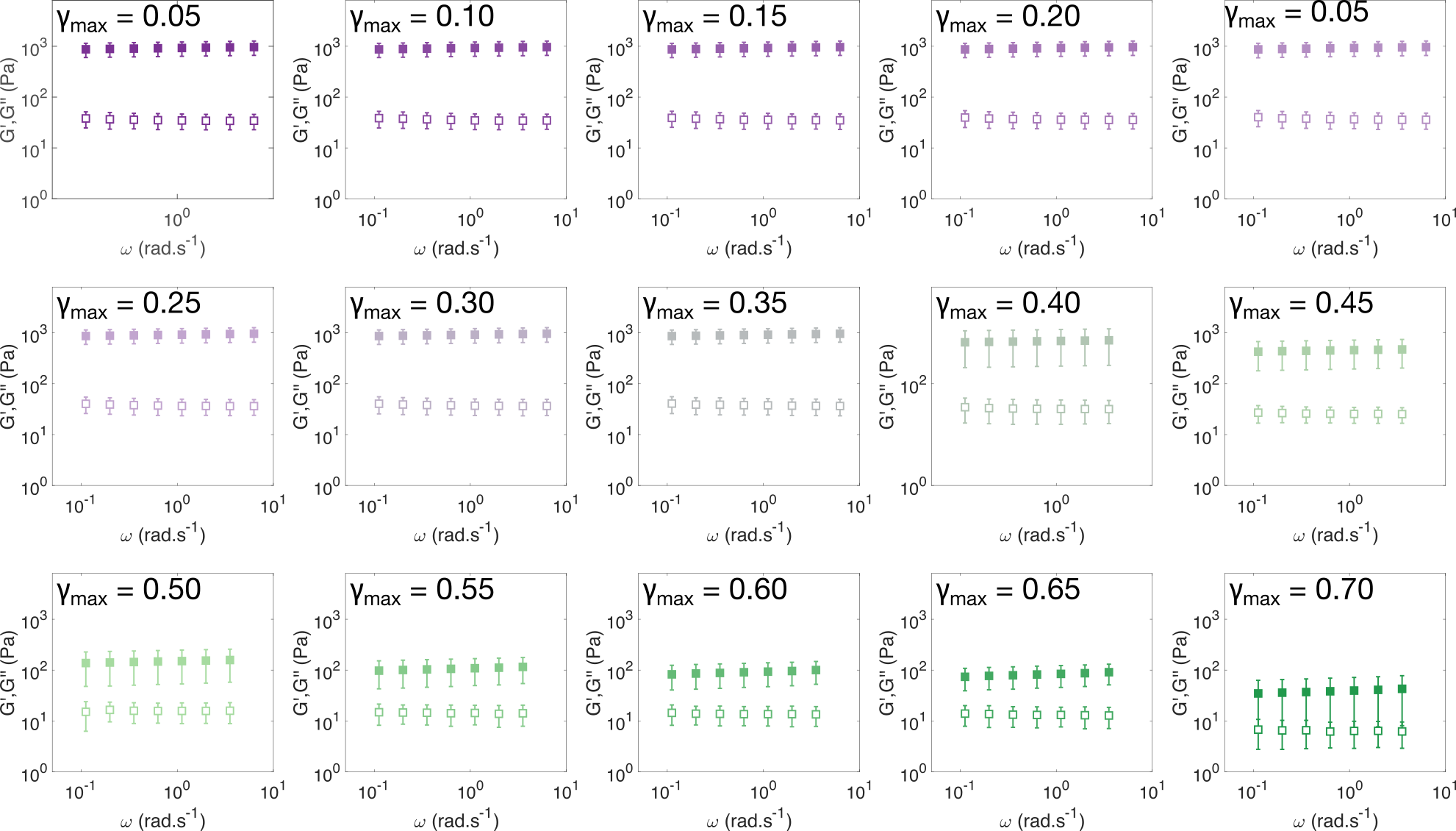}
    \caption{Representative frequency sweeps of the $I27_5$ hydrogels following the application of triangular strain pulses of increasing magnitude. Each frequency sweep probes the linear viscoelastic properties of the hydrogel after deformation. The spectra remain remarkably consistent at applied strains $\gamma$ below approximately 0.4, at which point the network undergoes irreversible fracture}
    \label{Supp - postloadingfreqsweeps}
\end{figure*}

The polyprotein structure comprises a concatenated series of five $I27$ protein domains as shown in figure \ref{supp-proteindesign}.  Linkers comprising 4-6 amino acids were inserted between domains to decrease inter-domain interactions. A hexahistidine tag is found at the N-terminus and a dicysteine group is found at the C-terminus. 

Dimensions and distances  are determined using the 'distance' tool in the molecular visualization software PyMOL. The individual $I27$ domains each have an N-C end-to-end length of $l_{domain}$ = 4.66 nm.  Each of the four linker domains contain between 4-6 amino acids with no intrinsic secondary structure.  Given that the average linker length is 5 residues, that the average length of an amino acid is 0.35 nm, and assuming that the linker adopts a freely-jointed-chain structure, the total average length of the linker $l_{link}$    is: $l_{link} = $0.35 nm$ \times \sqrt{5}   = $0.78 nm.

From this we may estimate the following:
\begin{enumerate}
    \item The mean distance between tyrosine crosslink sites is $l_{domain}+l_{link} = $5.44 nm
    \item The ratio between the linker length and polyprotein radius as $\frac{2l_{link}}{l_{domain}} = 0.34$. 
\end{enumerate}

\end{document}